\documentclass[%
 aip,
 amsmath,amssymb,
reprint,
fleqn
]{revtex4-1}

\usepackage{graphicx}
\usepackage{dcolumn}
\usepackage{bm}

\usepackage[utf8]{inputenc}
\usepackage[T1]{fontenc}
\usepackage{mathptmx}
\usepackage{etoolbox}
\usepackage{xcolor}

\makeatletter
\def\@email#1#2{%
 \endgroup
 \patchcmd{\titleblock@produce}
  {\frontmatter@RRAPformat}
  {\frontmatter@RRAPformat{\produce@RRAP{*#1\href{mailto:#2}{#2}}}\frontmatter@RRAPformat}
  {}{}
}%
\makeatother
\begin{document}

\preprint{AIP/123-QED}

\title[]{Clarifying the impact of dual optical feedback on semiconductor lasers through analysis of the effective feedback phase}
\author{Robbe de Mey}
\email{robbe.de.mey@vub.be}
  
\author{Spencer W. Jolly}%
 \altaffiliation[Also at ]{Service OPERA-Photonique, Université libre de Bruxelles (ULB), Brussels, Belgium.}

\author{Martin Virte}
\affiliation{%
Brussels Photonics (B-PHOT), Department of Applied Physics and Photonics, Vrije Universiteit Brussel, Pleinlaan 2, 1050 Brussels, Belgium
}%

\date{\today}

\begin{abstract}
Time-delayed optical feedback is known to trigger a wide variety of complex dynamical behavior in semiconductor lasers. Adding a second optical feedback loop is naturally expected to further increase the complexity of the system and its dynamics, but due to interference between the two feedback arms it was also quickly identified as a way to improve the laser stability. While these two aspects have already been investigated, the influence of the feedback phases, i.e. sub-wavelength changes in the mirror positions, on the laser behavior still remains to be thoroughly studied, despite indications that this parameter could have a significant impact. 
Here, we analyze the effect of the feedback phase on the laser stability in a dual-feedback configuration. We show an increased sensitivity of the laser system to feedback phase changes when two feedback loops are present, and clarify the interplay between the frequency shift induced by the feedback and the interferometric effect between the two feedback arms.
\end{abstract}

\maketitle

\begin{quotation}
We discuss the influence of the feedback phase on the dynamics of a semiconductor laser subject to two time-delayed optical feedback. While these phases are often neglected\cite{Ruiz-Oliveras2006} or considered as a perturbation on the delay \cite{Tobbens2008}, we highlight the large sensitivity of the laser behavior to these parameters. We discuss the effect of phase variations on the stability of the laser and its noise properties. We show that the frequency shift induced by the feedback plays an important role as well, and introduce the effective feedback phase difference as a key indicator capturing both the impact of the feedback phases and variations of the laser frequency.
\end{quotation}

\section{Introduction}
Semiconductor lasers subject to one optical feedback have been extensively studied for their dynamical behavior \cite{Ohtsubo2013,Uchida2012}. A wide range of feedback induced dynamics are possible: linewidth narrowing or broadening, periodic and quasiperiodic behavior, low frequency fluctuations or fully developed chaos in the output intensity, to name a few \cite{Tkach1986, Donati2013, Ohtsubo2013, Uchida2012}. Adding another optical feedback, thereby subjecting the laser to two distinct optical delays, further increases the complexity. The additional feedback has been shown to either stabilize or further destabilize the laser depending on the feedback parameters \cite{Rogister1999,Tavakoli2020,Liu1997,Ruiz-Oliveras2006}, to suppress chaotic dynamics \cite{Liu1996}, to significantly reduce the laser linewidth \cite{Lan2022,Zhang2022}, and to make the chaos more irregular\cite{Bakry2016}. Towards applications, it could be used for reservoir computing system \cite{Hou2018}, chaos control \cite{Rogister1999,Rogister2000}, or secure communication  \cite{Wu2009,Tronciu2008,Tronciu2008a,Lee2005}. This system has also been considered for improving self-mixing interferomety \cite{Zhu2018, Ruan2019, Mezzapesa2014}. Moreover, in photonic integrated circuits there is no practical optical isolator, hence on-chip lasers could easily be subject to multiple parasitic reflections potentially leading to undesired dynamical behavior. 

By adding a second optical feedback loop, the laser is, in a sense, coupled to an interferometer. The main difference with a mirror is that the reflection spectrum will then depend on the wavelength. In addition, it adds a dependence on the phase difference between the two feedback arms as such change will effectively detune the inteferometer and shift its reflection spectrum. It is however not obvious how this would impact the laser dynamics with a perfect laser. In a steady-state, the optical feedback systems acts exactly like an interferometer as long as the difference in the length of the arms is less than the coherence length of the laser (i.e. a few meters). But, as the laser is perturbed, this becomes inaccurate.
While the dual optical feedback case has attracted some attention, research on the impact of the feedback phase has been limited, even though it already tends to show that the phase impact can be significantly larger than in the single feedback case. As shown in Ref. \onlinecite{Lee2004}, bifurcation diagrams as a function of the gain can strongly differ if one of the feedback phases is changed. In Ref. \onlinecite{Tronciu2008}, the authors show that the number of stationary states (also called external cavity modes (ECMs)) and their exact values are depending on the value of the feedback phases. Moreover, they show that the number of modes and their specific solutions are significantly more complicated than the one-delay case. 

For some applications, the feedback phase plays a role \cite{Tronciu2008a,Hou2018,Rogister1999,Tobbens2008}. In Ref. \onlinecite{Tobbens2008}, the authors highlight that in a two-delay system without feedback phase control, the laser can fluctuate between stable and unstable behavior. Similarly, the time dependence of the intensity depends sensitively on the delay phase difference\cite{Fischer1994}, though a small change in one of the delays was not studied. Recently, in Ref. \onlinecite{FarBrusatori2022}, the authors have investigated the case of a laser with feedback from both sides of the cavity and show that with active control of the feedback phase they obtain a feedback insensitive lasing frequency. However, how the characteristics of the additional feedback changes the laser stability is unclear. In ref. \onlinecite{Barbosa2019}, studying a laser coupled to two optical feedback loops, the authors state, that the changes they see are not a result of optical frequency interferometry once the difference between feedback loops is long enough. In addition, in Ref. \onlinecite{DeMey2022}, we recently showed with others that in a two-delay system the feedback phases play a significant role in both the threshold current reduction and the time-delay signature arising in the chaotic output. In the end, to the best of our knowledge, in most work dealing with dual-optical feedback, the feedback phases were typically kept fixed, only changed implicitly via large delay changes, or not controlled directly \cite{Lee2005,Liu1996,Liu1997,Ruiz-Oliveras2006,Wu2009,Bakry2016,Rogister2000}.

In short, two points seem to be missing in the current understanding of a semiconductor laser coupled to double optical feedback. First of all, how does the feedback rate at which the Hopf bifurcation occurs, where the laser first becomes unstable, change due to the second delay? Secondly, how do the feedback phases influence the laser stability? 
In this paper, we investigate theoretically the stability of a semiconductor laser with double optical feedback with a focus on the effect of the feedback phases. In the first section, we introduce the system and the rate equations used for the simulations. 
Then we highlight that the second feedback loop can largely improve the stability of the laser, but only for specific combinations of time-delay and feedback phases for both feedback loops.
Next, we consider the effective feedback phase - which includes the variation of the laser frequency induced by feedback - and show that it is, in practice, a good indicator of the expected impact of the dual optical feedback on the laser behavior.
Finally, we show that below the first Hopf bifurcation the feedback phases impact the laser stability characteristics as measured by the Relative Intensity Noise (RIN).

\section{Simulation Model}

\begin{figure}
	\centering
	\includegraphics[width =0.66\linewidth]{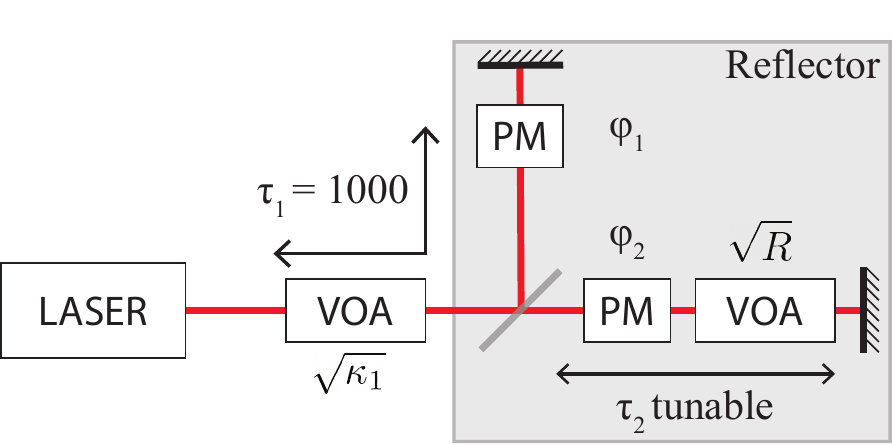}
	\caption{Schematic of the simulated system. The two variable optical attenuator (VOA) control the feedback rate $\kappa_1$ and $\kappa_2 = R\times\kappa_1$, while $\tau_1$ and $\tau_2$ correspond to the time-delay. The feedback phases $\phi_1$ and $\phi_2$ are independently set as if two independent phase modulators (PM) were used. The gray area indicates the reflector, composed of two mirrors at different positions, and of which the feedback rate ratio is fixed by R.}
	\label{fig:schematic}
\end{figure}

In this work, we take the following point of view: we consider that the two feedback loops are part of a single "black-box" reflector as shown in Fig. \ref{fig:schematic}. The idea is that all parameters inside this box would be fixed as if the reflector was an independent component e.g. like a microcavity with fixed strength, phase and time-delay differences between the two feedback loops. In particular, we use a fixed ratio between the feedback rates for the two loops as described below.

Numerically, we use an extended version of the Lang-Kobayashi (LK) equations \cite{Lang1980} including two optical feedback terms, normalized by the photon lifetime, similar to the ones shown in Ref. \onlinecite{Rogister1999}:
\begin{align}
\dot{E}(t) = &(1+i\alpha)N(t)E(t)+\kappa{}_{1}E(t-\tau_1)\exp(i\phi_1) \\ 
        &+\kappa{}_{2} E(t-\tau_2)\exp(i\phi_2 ), \nonumber \\ 
\dot{N}(t) = &(P-N(t)-(1+2N(t))E^2(t))/T
\end{align}

\begin{table}
\centering
	\begin{tabular}{ l l l }
		Symbol & Represents & Typical value/range \\ \hline
		$\kappa_1$ & Feedback rate feedback loop 1 & $0.1-5 \times 10^{-3}$\\
		$\tau_1$ & Delay feedback loop 1 & 1000 \\ 
		$\phi_1$ & Feedback phase feedback loop 1 & 0--2$\pi$ \\   
		$\kappa_2$ & Feedback rate feedback loop 2 & $R\times\kappa_1$\\
		$\tau_2$ & Delay feedback loop 2 & 50--	5000 \\  
		$\phi_2$ & Feedback phase feedback loop 2 &  0--$2\pi$   \\
		$R = \kappa_2/\kappa_1$ & Feedback rate ratio & 0--1\\
		$\alpha$  & Linewidth enhancement factor & 3 \\  
		P  & Pump parameter & 1 \\
		T & Carrier lifetime & $10^{3}$   
	\end{tabular}
	\caption{Details of the simulation parameters. We only change the feedback parameters ($\kappa_1$, $\phi_1$, $\kappa_2$, $\tau_2$, $\phi_2$). The laser parameters are fixed to these values.}
	\label{tab:LK}
\end{table}
\noindent where $E$ represents the complex electrical field, and $N$ represents the carrier density in the laser cavity. We fix the two feedback phases, $\phi_1$ and $\phi_2$, independently from the respective time-delay of each feedback loop. Indeed, from a physical point of view, the feedback phase would be equal to $\omega \tau$, with $\omega$ the lasing frequency and $\tau$ the time-delay. However, the two quantities can be rather independently adjusted considering the scaling differences: variations of the mirror position at the micrometer scale and below corresponds to phase variations, while changes at the centimeter scale and beyond would correspond to time-delay variations. In an experimental setup, this could be implemented as a linear translation stage to change the large-scale delay, together with a phase modulator or a piezo-actuator to change the delay on a sub-wavelength scale (see for example Ref. \onlinecite{DeMey2022}).

The feedback rate parameters are $\kappa_1$ and $\kappa_2$, though we will not set $\kappa_2$ directly. Instead, we will use the feedback rate ratio $R = \kappa_2 / \kappa_1$. $\alpha$, $T$ and $P$ are the linewidth enhancement factor, the carrier to photon lifetime ratio and the pump parameter respectively. $P = 1$ corresponds to two times the laser threshold while $P=0$ is the laser threshold. Apart from the last section, no spontaneous emission noise is considered.\\
For simplicity, we keep the first delay fixed at $\tau_1 =  1000$, and vary the second delay $\tau_2$. All parameter details can be found in table \ref{tab:LK}. Due to the normalization in time with respect to the photon lifetime, all time related parameters are dimensionless. To provide some insight for comparison purposes, with a photon lifetime of 3 ps, $\tau_1 = 1000$ corresponds to a delay of $\tau'  = 3000$ ps, i.e. a cavity of approximately $l' = c \tau'/2 = 45$ cm in free space.

Analytically, from the steady state solutions of the normalized equations without feedback ($\kappa_1 = \kappa_2 = 0$), we can calculate the Relaxation Oscillation Frequency (ROF) of the solitary laser. Following the approach described in Ref. \onlinecite{Ohtsubo2013}, the ROF is: $v_R = \frac{1}{2\pi}\sqrt{\frac{2P}{T}-\frac{1}{4}(\frac{2P+1}{T})^2}$. With the parameters used in this work, we get a Relaxation Oscillation Period (ROP): $\tau_{ROP} = 1/v_R = 141$. For most simulation discussed in this work, $\tau_1$ and $\tau_2$ are set at longer delays than $\tau_{RO}$ which typically places us in the long cavity regime. 

\section{Phase-dependent stabilization enhancement triggered by the second feedback loop}
In this section, we study the stability of the laser with the feedback rate as bifurcation parameter when varying the second delay and feedback phases configuration. We identify the first Hopf bifurcation occurring for increasing feedback rates based on the number of extrema recorded in the simulated time-series.

\begin{figure}
	\centering
	\includegraphics[width = \linewidth]{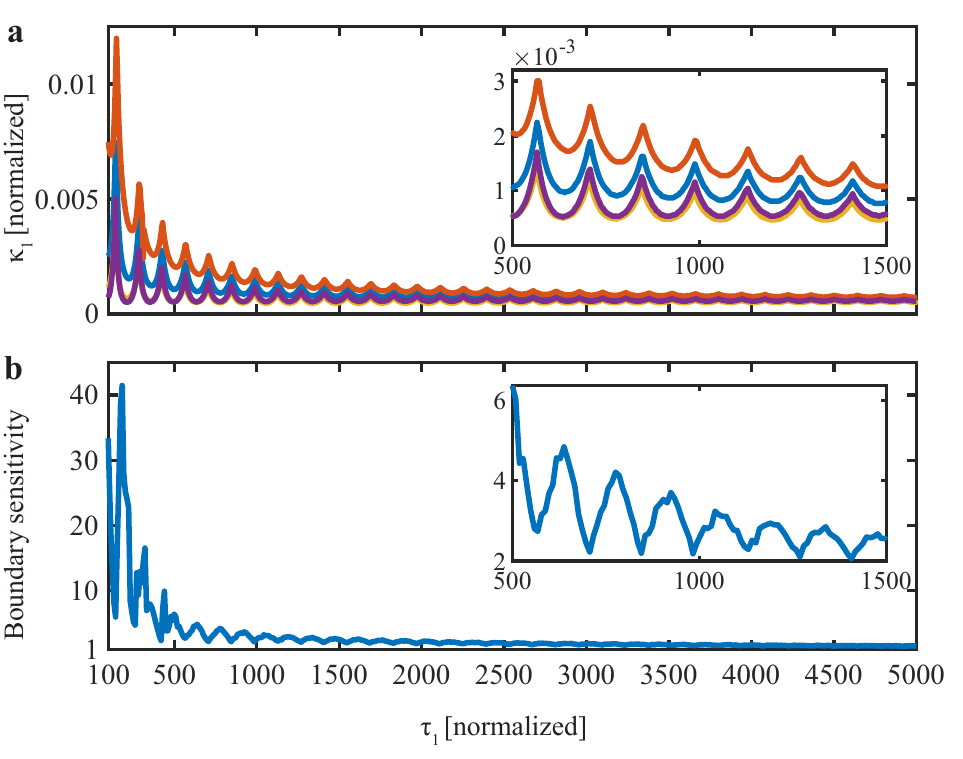}
	\caption{\textbf{a} Stability boundary of a laser with one-delay for different feedback phase ($\phi_1$) values: $0$ (blue), $\frac{\pi}{2}$ (orange), $\pi$ (yellow), $\frac{3\pi}{2}$ (purple). Each curve indicates the transition from a stable steady-state (below) to unstable behavior (above). \textbf{b} Sensitivity of the stability boundary against the feedback phase as a function of the time-delay. As described in the text, a high value indicates a large change of the stability boundary position when the feedback phases are changed.}
	\label{fig:oneDelay}
\end{figure}

To have a baseline, we first consider the case of a laser with a single time-delayed feedback. In this case, as shown in Fig. \ref{fig:oneDelay} a, the position of the Hopf bifurcation---corresponding to the boundary below which the laser stays in a stable state---with respect to the delay of the feedback cavity exhibits multiple peaks of increased stability which relate to the ROF of the laser \cite{Fan2015}. This effect gradually decreases as the delay is increased. In Fig. \ref{fig:oneDelay} a, we show the stability boundaries for four different feedback phase values. We can thus observe that changing the feedback phase for short delays strongly impacts the stability boundary. On the other hand, in the case of long delays, the stability boundary seems to asymptotically converge to a single value independent of the delay. However, at $\tau_1 = 5000$ the boundary sensitivity is still 1.5, indicating that the feedback phase still has an impact on the boundary although the impact is much less than for short delays. This illustrates that, as already known, the impact of the feedback phase is significantly smaller in the long cavity regime. As a comparison for later simulations, for the system with only one-delay at $\tau_1 = 1000$, $\phi_1 = 0$, and $R = 0$, the laser becomes unstable for $\kappa_1 > 1.26 \times 10^{-3}$. We further highlight the impact of the feedback phase by estimating the ratio between the largest and smallest feedback rates at which the first Hopf bifurcation occurs across all phase values. In a sense, it gives a measure of the sensitivity of the stability boundary to the feedback phase. For a given delay, we vary the feedback phase over a full period ($0$ to $2\pi$) and compute for each phase value the position of the Hopf bifurcation. By dividing the maximum and minimum feedback rate found, we calculate the "boundary sensitivity" ratio. The result is shown in Fig. \ref{fig:oneDelay} b as a function of the time-delay $\tau_1$. At this stage, we clearly observe that the feedback phase has an impact on the stability beyond the short cavity regime, although it gradually reduces for longer delays. For $\tau_1 = 1000$ the boundary sensitivity is $2.58$, while it is reduced to $1.39$ for $\tau_1 = 10000$. These values however remain well below the sensitivity recorded for shorter cavities which peaks above $40$.

\begin{figure}
	\centering
	\includegraphics[width = \linewidth]{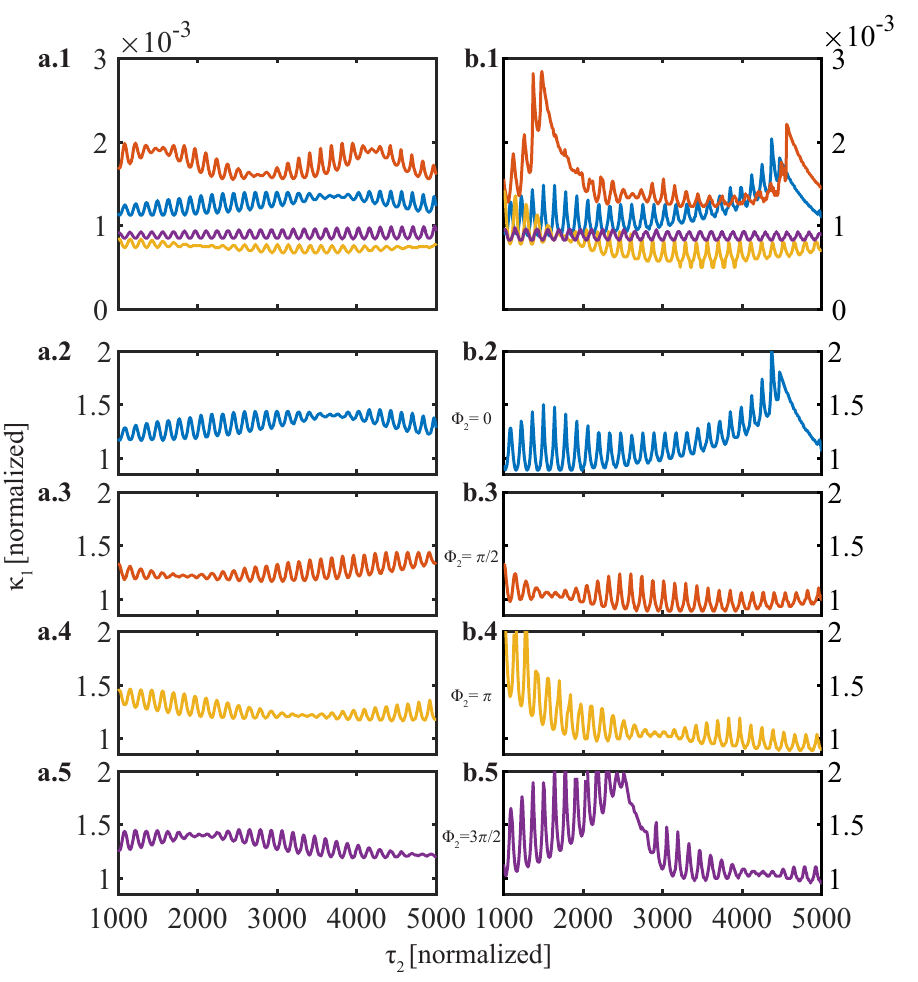}
	\caption{Stability boundaries with respect to $\tau_2$ for different values of the feedback phases, comparing $R = 0.1$ (\textbf{a}) with $R = 0.4$ (\textbf{b}). \textbf{a.1} $\phi_2 = 0$, and different $\phi_1$ values: $0$ (blue), $\frac{\pi}{2}$ (orange), $\pi$ (yellow), $\frac{3\pi}{2}$ (purple).  \textbf{a.2}-\textbf{a.5} $\phi_1 = 0$ and different $\phi_2$ values (given on right side of each panel). \textbf{b.1}-\textbf{b.5} sweeping the same feedback phase parameters but for $R = 0.4$.}
	\label{fig:firstHopf} 
\end{figure}

We now fix the first time-delay at $\tau_1 = 1000$, and turn on the second feedback loop by tuning the feedback ratio $R$ upwards from zero. We then perform the same simulations as for the single-delay case, but now changing the second time-delay $\tau_2$. We observe that adding a second feedback loop impacts the stability boundary significantly as $R$ is increased, as shown in Fig. \ref{fig:firstHopf}. 
At $R = 0.1$ significant differences in the stability boundaries appear. In Fig. \ref{fig:firstHopf} a.1 the peaks of increased stability related to the ROF are still visible. However, we observe additional bumps of stability at certain $\tau_2$ values. In contrast to the one-delay case, these new peaks depend strongly on the feedback phases. As can be seen by the stark difference between the lines in Fig. \ref{fig:firstHopf} a.1, corresponding to different values of $\phi{}_{1}$. By only changing $\phi_1$ the position of the stability boundary moves from $\kappa_1 \approx 0.8 \times 10^{-3}$ to $\kappa_1$ $\approx 1.8 \times 10^{-3}$, everything else remaining equal. Around a delay of 1000, this is similar to the one delay case (see the inset in Fig. \ref{fig:oneDelay} a). However, towards longer delays the stability boundaries do not converge that quickly anymore. Indeed, $\phi_1$ has a distinct impact on the periodicity of stability bumps with respect to $\tau_2$. For most values of $\phi_1$, the period takes on high values, making the stability boundary almost flat for a large delay change. For $\phi_1 = \pi/2$ the large stability enhancements occur for every $\tau_2$ increment of 2500, making the stability of the laser strongly dependent on $\tau_2$. On the other hand, if we keep $\phi_1$ constant and sweep $\phi_2$, as shown in Fig. \ref{fig:firstHopf} a.2, the whole pattern of the stability boundary shifts horizontally with respect to $\tau_2$. 
When increasing the feedback ratio to $R = 0.4$, the enhanced stability regions are more pronounced, see Fig. \ref{fig:firstHopf} b.1 and b.2. Now tuning either $\phi_1$ or $\phi_2$ changes the peaks in the stability boundary or shifts it horizontally, or a combination of both. In addition, the added stability peaks can be sharper and the stability boundary shifts from $\kappa_1 \approx 0.5 \times 10^{-3}$ to $\kappa_1 \approx 2.8 \times 10^{-3}$.

The stabilization enhancements seem to exhibit some periodicity with respect to the second time-delay $\tau_2$. While the phase of the second - and weaker - optical feedback $\phi_2$ seems to simply shift the stability boundary along the $\tau_2$ axis, both the amplitude and the period of the stabilization enhancement appear to be impacted by a change of the feedback phase $\phi_1$ corresponding to the stronger feedback arm. These results suggest that changing feedback parameters not only modify the spectral (interferometric) effect of the dual-cavity configuration but also the laser emission itself.

\begin{figure}
	\centering
	\includegraphics[width = \linewidth]{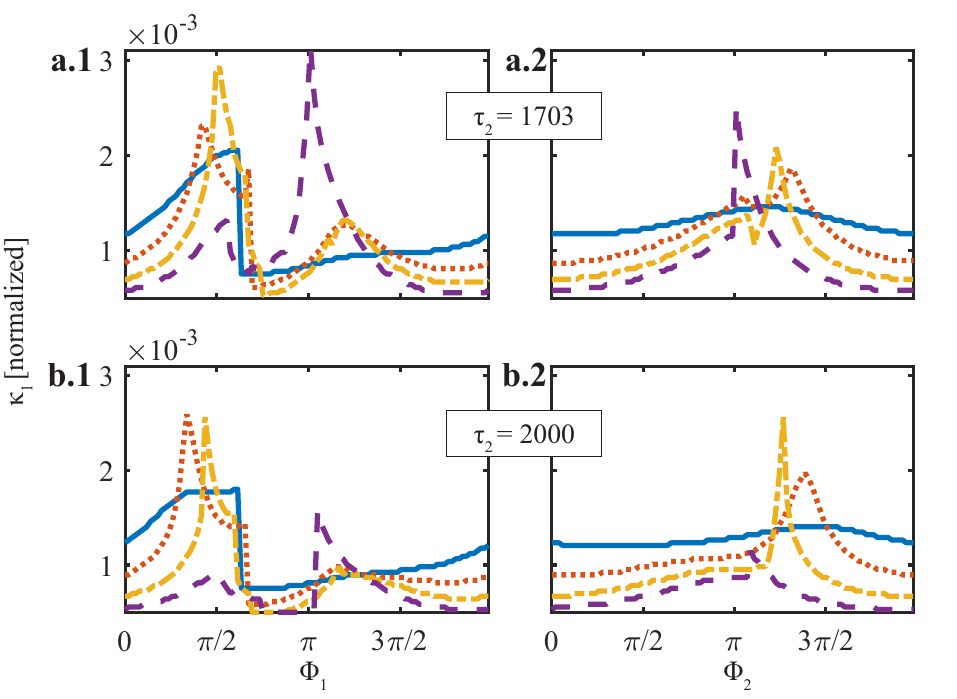}
	\caption{Stability boundaries with respect to the feedback phases for different R values: R = 0.1 (blue solid), 0.4 (orange dotted), 0.7 (yellow dash-dotted), 1 (purple dashed). \textbf{a} $\tau_2 = 1703$, \textbf{b} $\tau_2 = 2000$. \textbf{a.1} and \textbf{b.1} $\phi_2 = 0$. \textbf{a.2} and \textbf{b.2} $\phi_1 = 0$.}
	\label{fig:boundaries_phases_sweep}
\end{figure}

Previous results also indicate that the R value plays an important role in the emergence of the enhanced stability region. This is logical since complete destructive interference between the two feedback arms would only be achievable for $R=1$. To find the dependence of the feedback phase on this R parameter, we now keep $\tau_2$ fixed and sweep the feedback phases for different R values. We consider two different delays: $\tau_2 = 1703$ and $\tau_2 = 2000$. $\tau_2 = 1703$ is chosen so that the greatest common denominator between the two time-delays is 1, else the feedback cavity might act as the one-feedback case \cite{Lee2005}. For each configuration, we fix either $\phi_1$ or $\phi_2$ to 0 and sweep the other phase term from $0$ to $2\pi$. We perform these simulation for $R$ equal to $0.1$, $0.4$, $0.7$ and $1$. The results are plotted in Fig. \ref{fig:boundaries_phases_sweep}. 
For very low R values, $R = 0.01$ (not shown), the second delay is almost off, and the stability boundaries only show a change for sweeping $\phi_1$. Even for $R = 0.1$ the impact is limited, although some change due to the feedback phase is clearly visible in the case of tuning $\phi_2$ in panel a.2 and b.2. When increasing R further, we get a mix of the effect of sweeping $\phi_1$ and a new peak due to the destructive interference. The latter is very obvious for $R=1$ (purple dashed curve) especially for $\phi_1=\pi$ and $\phi_2 = 0$ in panel a.1 and b.1. Interestingly this peak doesn't appear so prominently in the symmetric case with $\phi_1 = 0$ and $\phi_2 = \pi$ (a.2 and b.2) which suggest an influence of the time-delay as well, i.e. a deviation from what we expect from a simple interferometer. In the same case of $R=1$, we see that the discontinuity observed when sweeping $\phi_1$ is largely suppressed and the peak corresponding to destructive interference is the dominant feature. However, for the intermediate cases of $R=0.4$ and $R=0.7$, we observe more complex oscillations around the pattern of the single feedback case: the discontinuity is still the dominant feature, but the amplitude of the variations are increased with a notable peak below $\pi/2$ when sweeping $\phi_1$. The peak corresponding to destructive interference seem to appear for $\phi_1$ values larger than $\pi$ and to gradually shift and grow in amplitude. When sweeping $\phi_2$, we observe one dominant peak that again seems to gradually grow in amplitude and shift from values above $\pi$ to $\phi_2=\pi$ for $R=1$. In the case of $\tau_2 = 2000$, we remark that this is increasingly sharp when going from $R = 0.4$ to $R = 0.7$. However going from $R=0.7$ to $R=1$ leads to a significant reduction of the peak. The origin of this feature is however unclear.

As mentioned earlier, for these simulations, we start with a feedback rate well below the stability boundary and increase it gradually until the simulated intensity time-series exhibit oscillations or instabilities, which then give us the stability boundary. However, by doing so, we always reset the history of the system when changing the feedback phase. In this way, we can not check if the system has hysteresis when changing these parameters. To check for hysteresis around the sharp edges, we can fix the feedback rate and sweep the feedback phase ($\phi_1$) from 0 to $2\pi$ and back to 0. With this approach, we confirm that, as in the single feedback case, a broad hysteresis can be found around the discontinuity of the stability boundary. In the case of a large R, however, the situation appears to be more complex as multiple regions of hysteresis occur. But this investigation is out of the scope of this paper and is, for now, left for future work.

\begin{figure}
	\centering
	\includegraphics[width = \linewidth]{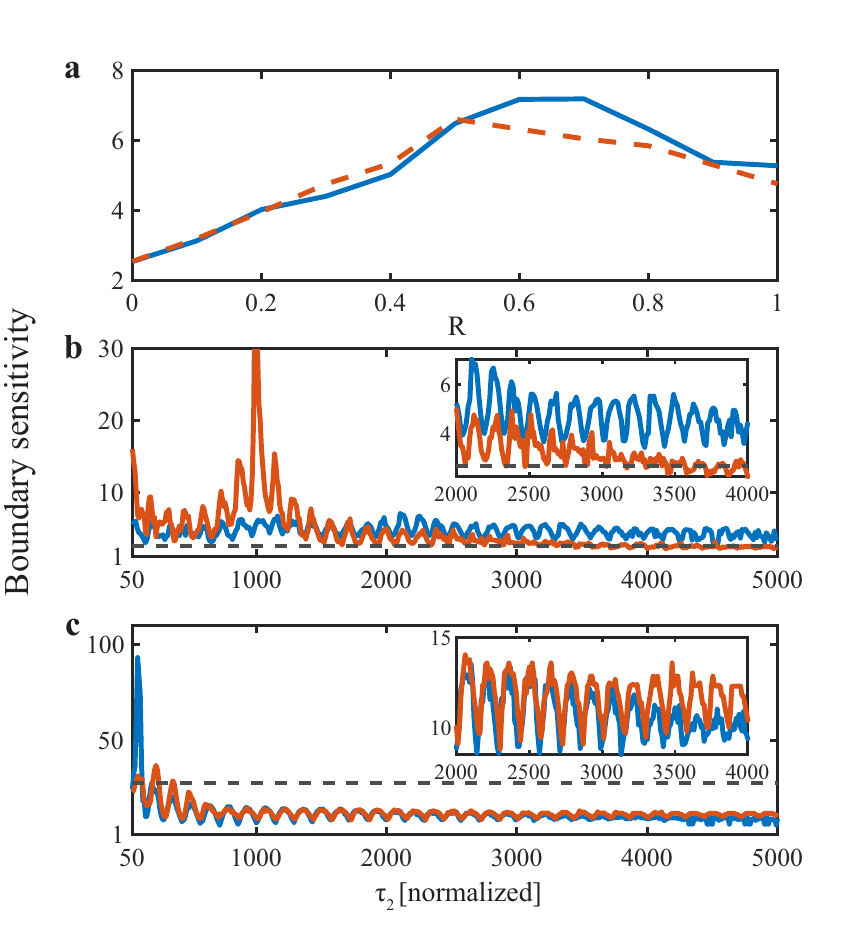}
	\caption{Impact of the feedback phases on the stability of the laser. The y-axis represents the sensitivity of the stability boundary on the feedback phases. \textbf{a} Sensitivity when changing R, for $\tau_2= 1703$ (blue) and $\tau_2 = 2000$ (orange). \textbf{b}  Sensitivity when changing $\tau_2$, with $R = 0.4$ (blue) and $R = 1$ (orange). The grey horizontal line indicates the sensitivity of the stability for $R =0$. \textbf{c}  Sensitivity when changing $\tau_2$, with $R = 0.4$ (blue) and $R = 1$ (orange) for $\tau_1 = 100$. The grey dashed horizontal line indicates the sensitivity of the stability boundary for $R =0$ and $\tau_1 = 100$.} 
	\label{fig:phaseImpact}
\end{figure}

It is now clear that the stability boundary, i.e. the position of the first Hopf bifurcation, will typically exhibit wider fluctuations when the feedback phase is varied in the dual-feedback configuration than in the single feedback case. To better quantify the relative importance of the feedback phases between the two configurations, we compute the sensitivity of this boundary to the feedback phases as done in the previous section for the single feedback system and shown in Fig. \ref{fig:oneDelay}. In practice, we use the same approach, though instead of sweeping only one phase term, we sweep the two phase terms independently. The results are shown in Fig. \ref{fig:phaseImpact}.
In panel a, we first analyze the effect of $R$, with $\tau_2 = 1703$ and $\tau_2 = 2000$ for the same reasons as stated earlier. Here $\tau_1$ is still equal to 1000. We observe that the maximum sensitivity is reached before $R=1$, that is for an unbalanced feedback strength between the two arms. Interestingly the maximal phase dependence is not achieved for the same ratios between our two cases. 
In panel b, we plot the evolution of the sensitivity as a function of $\tau_2$ for $R=0.4$ and $R=1$ in blue and red respectively, again with $\tau_1 = 1000$. For the unbalanced case, we observe that the sensitivity is above the single feedback case, but roughly similar for all values of $\tau_2$ considered in the plot. Oscillations matching the impact of the RO period as already highlighted before are visible but their amplitude - which is bigger than in the single-feedback case - seem to remain rather constant across the whole range of time-delays. For $R=1$, the situation is quite different: a strong peak occurs when $\tau_2$ is close to the value of $\tau_1 = 1000$, and a smaller peak emerges for short cavities $\tau_2< 100$. The large central peak is, of course, expected as this configuration simply approaches the perfect interferometer configuration and thus the ability to have fully destructive interference which would, in effect, suppress completely the amount of light sent back towards the laser. For longer delays, the sensitivity seems to asymptotically tend to the same value reported for the single-feedback case and shown as a dashed gray line. 
Last, in panel c, we show the evolution of the sensitivity as a function of $\tau_2$ but for a first feedback in the short cavity regime $\tau_1 = 100 < \tau_{ROP} \approx 141$. We consider again feedback ratios of $R=0.4$ and $1$ in blue and red respectively. We thus observe that apart from a large peak for $\tau_2<100$ and $R=0.4$, adding a second feedback loop significantly reduces the sensitivity of the stability boundary to the feedback phase. For larger time-delays $\tau_2$, there is no clear difference between the two feedback ratios. 

\begin{figure*}
	\centering
    	\includegraphics[width = \textwidth]{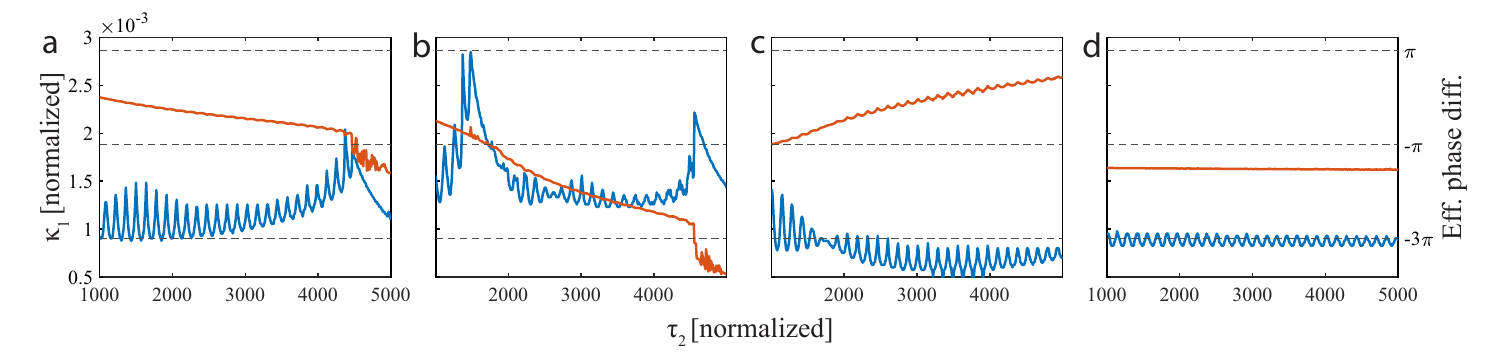}
	\caption{Stability boundaries (blue) and effective phase differences (orange) versus the time-delay $\tau_2$ for $R=0.4$ and $\phi_2 = 0$. From left to right: $\phi_1$ = $0$, $\frac{\pi}{2}$, $\pi$, $\frac{3\pi}{2}$.}
	\label{fig:effectivPhase}
\end{figure*}

These results show that there does not seem to be a clear short or long cavity regime in the case of double optical feedback. Instead, due to the interference between both feedback loops, and the interaction with the laser frequency, the stability boundary is sensitive to the feedback phases for a large delay range. In the one-delay case there is a clear impact of the feedback phase when the delay is in the short cavity, while the feedback phase can be partially neglected for a delay far into the long cavity regime. In comparison, the two-delay system is sensitive to the feedback phase for a much larger delay range. However, overall the sensitivity to a feedback phase change remains higher for a strong feedback close to the laser. The feedback phases should be considered for both small and large delay differences between both feedback loops. In contrast to the case for only one feedback loop, there does not seem to be a clearly defined delay for which the feedback phases can be neglected.\\

\section{Effective feedback phase and spectral dependencies}
In view of the previous results, the stabilization enhancement effect seems to be likely due to the combination of interferences between the two feedback arms and the wavelength shift induced by the feedback on the laser \cite{Ohtsubo2013}. If the lasing wavelength shifts, the phase with which light returns to the cavity is directly impacted. To reconcile the different points of view, we define the effective feedback phase, i.e., the feedback phase effectively experienced by the laser including the effect of a wavelength variation. For each feedback loop, the effective phase is calculated as: 
\begin{equation}
	\phi_\textrm{eff,i} = \tau_i\delta+\phi_i,
\end{equation}
\noindent in which $\delta$ is the change in the lasing wavelength from the free-running case, with $\tau_i$ the delay and $\phi_i$ the feedback phase of each arm. We estimate this frequency shift in our simulations by taking the average of the change of the phase of the electric field: $\delta = \langle\frac{d\phi(t)}{dt}\rangle$. At this point, it is crucial to mention that the two phase terms $\phi_1$ and $\phi_2$ are defined for the wavelength of the laser without feedback, and therefore do not vary if the laser wavelength shifts. Upon returning to the cavity, light from both arms interferes. By taking the difference between the effective feedback phase of feedback loop one and two, we can estimate when the feedback from each arm interferes constructively or destructively. The effective phase difference directly reads as: 
\begin{equation}
	\Delta\phi_\textrm{eff} = (\tau_2-\tau_1)\delta+\phi_2-\phi_1.
\end{equation}
To confirm that the effective phase difference fully captures the spectral and phase effects, we plot it together with the stability boundaries for the case of $R = 0.4$ in Fig. \ref{fig:effectivPhase}. When the effective phase difference is close to a multiple of $\pi$ which corresponds to a case of destructive interference between the two feedback arms, we indeed observe that the stability is significantly enhanced. The same observations can be made for the case of $R = 0.1$. 
The effective phase difference therefore seems to be an accurate indicator to predict the feedback configuration that would lead to enhanced stability. Yet, the value of $\delta$, measured just below the stability boundary, is crucial, but analytically determining the relation between $\delta$, i.e., the wavelength shift of the laser, and the feedback parameters is significantly more difficult than in the one delay case, mostly due to the feedback phase interaction in the feedback cavity for long delays. This involving calculation is, for now, left for future work. 

The effective phase shows where the additional stability occurs. To better understand what is happening we now look at the response of the feedback section. We calculate the response as follows:
\begin{equation}
	H  = \exp(-i(\delta \tau_1+\phi_1))+R\exp(-i(\delta \tau_)+\phi_2)).
\end{equation}

\noindent By taking the angle and real part of this complex function we show the response of the feedback section and the interaction with the laser. We shall study two cases, for $R = 0.4$ and $R = 0.7$. 

\begin{figure}
	\centering
    	\includegraphics[width = \linewidth]{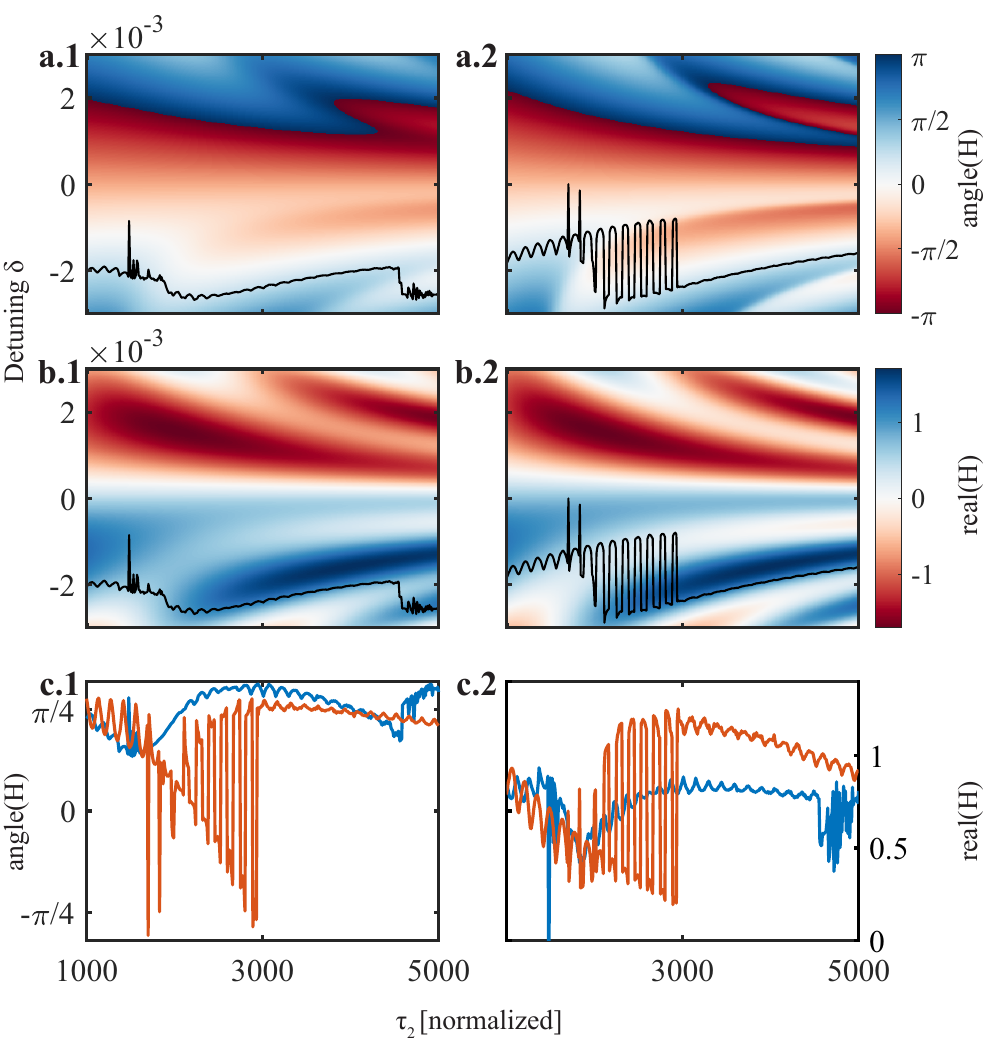}
	\caption{Response of the feedback section. \textbf{a.1}, \textbf{b.1}: $R = 0.4$. \textbf{a.2}, \textbf{b.2}: $R = 0.7$. Simulations parameters: $\tau_1 = 1000$, feedback phases are constant: $\phi_1 = \pi/2, \phi_2 = 0$. Black lines indicate the $\delta$ at the stability boundary as measured from the rate equation simulations. \textbf{c.1}, \textbf{c.2}: The angle and real part of the response along the black lines in the maps above for $R = 0.4$ (blue) and $R = 0.7$ (orange).}
	\label{fig:spencerMap}
\end{figure}

Figure \ref{fig:spencerMap} shows the response when we keep the feedback phases fixed while sweeping the detuning $\delta$ and the second delay $\tau_2$. a.1 and a.2 show the angle of the response, which shows that due to the second feedback arm the phase shift due to the feedback section can vary in a more complex way, which in turn leads to complex changes in $\delta$. Moreover, this phase shift depends strongly on the ratio of the feedback strength R, as seen in the difference between the a.1 ($R = 0.4$) and a.2 ($R = 0.7$). b.1 and b.2 show the real part of the response. The black lines in the plot shows $\delta$ at the stability boundary as measured from rate equation simulations (indirectly shown in Fig. \ref{fig:effectivPhase} through the effective phase for $R = 0.4$). In addition, b.1 and b.2 show the real part of the response. A local minimum in the real part of the response seems to predict the peaks in the stability boundary. In this sense it works as an extension of predicting the enhanced regions of stability with the effective phase difference described above. Local minimima in the real part of the response (along the black line) also show the small regions of increased stability linked to the relaxation oscillation period. The angle and real part along the black lines is plotted in c.1 and c.2. For both cases the phase value remains approximately fixed to 0.7 along the stability boundary, although clearly for $R = 0.7$ strong oscillations appear. These oscillations occur for $\tau_2$ between 2000 and 3000.  

\section{Effect of feedback phase variations on the Relative Intensity noise}
Finally, we show that the feedback phases also impacts the laser characteristics even when the laser is stable. A simple metric to represent this effect is the Relative Intensity Noise (RIN). A lower RIN can improve the laser characteristics e.g. through a reduction of the linewidth which is beneficial for many applications including telecommunication or sensing. To analyze the RIN of the laser in our simulation, we add a spontaneous emission noise term to our equations with a noise parameter $\beta = 10^{-9}$. We can then directly calculate the variance of the intensity and divide it by its mean value and obtain the RIN value. 

\begin{figure}
	\centering
	\includegraphics[width = \linewidth]{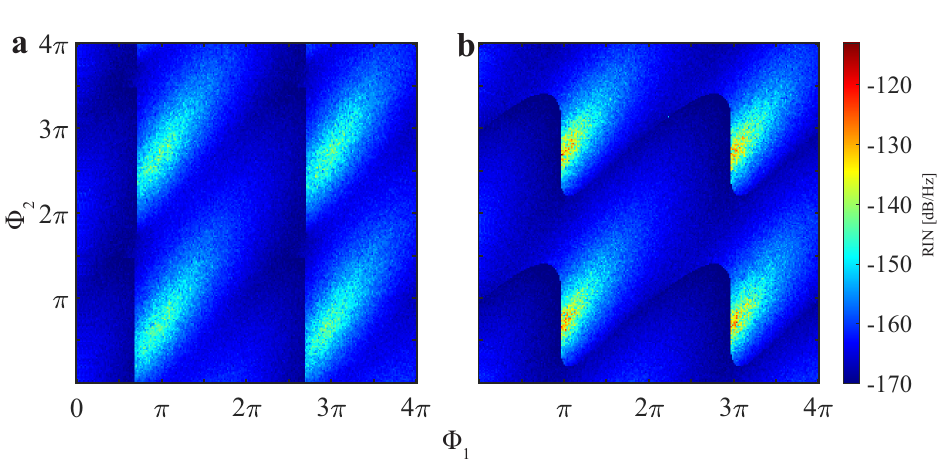}
	\caption{Impact of the feedback phases on the RIN of the laser when in a stable steady-state. \textbf{a} $\kappa_1 = 0.5 \times 10^{-3}$, $R = 0.4$. \textbf{b} $\kappa_1 = 0.35 \times 10^{-3}$, and $R = 1$. $\kappa_1$ adjusted so the total feedback rate is the same in \textbf{a} as in \textbf{b}. The time-delays are fixed in both cases to $\tau_1 = 1000$, and $\tau_2= 1703$.} 
	\label{fig:RIN}
\end{figure}

In Fig. \ref{fig:RIN}, we show the evolution of the RIN when sweeping the two feedback phases over two complete periods, from $0$ to $4\pi$, with $\tau_1 = 1000$ and $\tau_2 = 1703$ in the case of $R = 0.4$ (a) and $R = 1$ (b). To make a comparison we adapt $\kappa_1$ such that the total feedback ($\kappa = \kappa_1+\kappa_2 = \kappa_1(1+R)$) remains constant. For these simulation the laser is always in a stable steady-state, i.e. at a feedback rate below the stability boundary.
We observe that the RIN clearly depends on both feedback phases with some discontinuities. For $R = 0.4$ and $\phi_1 \approx 2.1$, as shown in Fig. \ref{fig:RIN} a, we observe a clear jump consistent with the laser switching to another external cavity mode with a different frequency. The minimum RIN is $-172\, dB/Hz$, the maximum is $-136\, dB/Hz$. On the other hand, the evolution observed for $R=1$ appears to be smoother but the occurrence of discontinuities still appears, although now when sweeping either feedback phase. For this case the minimum RIN is $-173\, dB/Hz$, similar to the previous case. The maximum RIN is $-114\, dB/Hz$, clearly higher. In short, even when working with the laser in a stable state, a small sub-wavelength change of the mirror position can impact the lasing properties. 
It is interesting here to notice the analogy with the case of a two-state emitting quantum dot laser but with a single optical feedback loop \cite{Pawlus2017}. Similar patterns are observed as each emitting mode experiences a different feedback phase due to the wavelength difference, hence leading to a synthetic dual-feedback configuration. 

To compare with the one-delay case we did a simulation for R = 0, $\kappa_1 = 0.7\times 10^{-3}$, and $\tau_1 = 1000$. By sweeping $\phi$ from $0$ to $2\pi$ the RIN changes from $-170 \, dB/Hz$ to $-146 \, dB/Hz$. So the minimal RIN value recorded in the simulation of Fig. \ref{fig:RIN} is similar to the one-delay case. However, in the two-delay case, the RIN can reach higher values depending on the feedback phase values. From these results, we could assume that undesired or parasitic back reflections could lead to significant degradation of the noise levels in the laser. Yet, recent experimental efforts report large reduction of the laser linewidth in a similar two-feedback configuration \cite{FarBrusatori2022}, thus suggesting that the noise impact could be strongly configuration dependent. 

\section{Conclusion}
In this work, we have numerically investigated the stability of a semiconductor laser subject to a dual-optical feedback and analyzed in more details the impact of the feedback phases. We showed clear qualitative differences on the laser behavior when an additional feedback loop is considered, and we point out the importance of the feedback phases on these effects. Due to interferences occurring in the feedback section and the effect of the feedback on the laser frequency, variations of any of the feedback phases can strongly modify the stability properties of the laser. 
We showed that the dependency of the stability on the feedback phases is strongest for intermediate values of the feedback ratio $R$, corresponding to unbalanced feedback strengths, and when both feedback loops corresponds to a long cavity. The most notable exception is when the two time-delays are close to each other; in this case, destructive interferences between the two feedback arms can drastically reduce the impact of the feedback on the laser and thus to a large enhancement of the stability. On the other hand, with $R < 0.1$, the impact of the second feedback is small, perhaps even negligible to some extent. 
If the dominant feedback loop is in the short cavity regime, adding a second feedback with a longer time-delay will typically reduce the influence of the feedback phase compared to the one-feedback case. The sensitivity however remains relatively high in this case. 
Moreover, we show that, while the effective feedback phase appears to be a useful indicator to deduce the potential impact of a given dual-feedback configuration on the laser, it remains difficult to derive one general formula between the feedback parameters that would lead to enhanced stability or higher complexity.
Finally, we have briefly investigated the impact of the dual-feedback configuration on the noise properties of the laser. Again we report a significant influence of the feedback phases and show that the RIN can be significantly increased.

Given that the enhanced stability is due to interferences in the feedback sections, it can be expected that similar phenomena persist for lasers coupled to more than two optical delays. Moreover, we also showed that even at $R = 0.1$, there is a clear impact on the laser stability. In a system with multiple parasitic reflections, for example on photonic integrated circuits, this could lead to problems. As the fabrication tolerances are typically in the nanometer scale, the laser characteristics could be strongly impacted by multiple parasitic reflections occurring on a chip. Further investigations might therefore be relevant to further improve the properties and reliability of lasers in photonic integrated circuits. 

\begin{acknowledgments}
The authors acknowledge funding from Fonds Wetenschappelijk Onderzoek (FWO), project G0E7719N. Vlaamse Overheid, METHUSALEM program, European Union H2020 research and innovation program, Marie Sklodowska-Curie Action (MSCA), project 801505. 
\end{acknowledgments}

\section*{Data Availability Statement}
The data that support the findings of this study are available from the corresponding author upon reasonable request.

\section{Conflict of Interest}

The authors have no conflicts to disclose.

\section{Author Contributions}

Robbe de Mey: formal analysis (equal); investigation; writing - original draft; writing - review and editing (equal). Spencer W. Jolly: formal analysis (equal); supervision (supporting); validation (equal); writing - review and editing (equal). Martin Virte: formal analysis (equal); methodology; supervision (lead); validation (equal); writing - review and editing (equal).

\section*{References}
\bibliography{aipsamp}

\end{document}